\documentclass[doublecol,figures]{epl2}

\usepackage{epsfig}
\usepackage{amsmath}
\usepackage{amsfonts}
\usepackage{amssymb}
\usepackage{graphicx}
\usepackage{units}
\usepackage{url}

\newcommand{\be}{\begin{equation}}
\newcommand{\ee}{\end{equation}}
\newcommand{\bea}{\begin{eqnarray}}
\newcommand{\eea}{\end{eqnarray}}

\DeclareMathOperator\atan{atan}
\DeclareMathOperator\tanhyper{th}
\DeclareMathOperator\im{Im}

\title{The third virial coefficient of a two-component unitary Fermi gas across an Efimov-effect threshold}
\shorttitle{The third virial coefficient of a two-component unitary Fermi gas across an Efimov-effect threshold}
\author{Chao Gao$^*$, Shimpei Endo and Yvan Castin}
\shortauthor{Chao Gao, Shimpei Endo and Yvan Castin}
\institute{$^*$Institute for Advanced Study, Tsinghua University, Beijing, 100084, China \\
Laboratoire Kastler Brossel, ENS-PSL, CNRS, UPMC-Sorbonne Universit\'es, Coll\`ege de France, Paris, France}


\abstract{We consider a mixture of two single-spin-state fermions with an interaction of negligible range and infinite $s$-wave scattering length.
By varying the mass ratio $\alpha$ across $\alpha_c\simeq 13.6069$ one can switch on-and-off the Efimov effect.
We determine analytically the third cluster coefficient of the gas. We show that it is
a smooth function of $\alpha$  across $\alpha_c$ since, unexpectedly, the three-body parameter characterizing the interaction
is relevant even on the non-Efimovian side $\alpha<\alpha_c$.}

\begin{document}
\date{15 January 2015}
\maketitle

\section{Introduction}

A powerful theory tool in the statistical physics of interacting quantum systems is the so-called cluster or virial
expansion, where the thermodynamic potentials are expanded in powers of the small degeneracy parameter
\cite{Huang}. Whereas the second cluster coefficient $b_2$ had a known general expression since
the 1930s \cite{BethUl}, it has been a long-lasting challenge to determine the third cluster coefficient $b_3$ explicitly.
Starting from the late 1950s,
analytical results for $b_3$ have been obtained for the two-body 
hard-core model, the archetype of non-resonant interactions where the $s$-wave scattering length $a$ 
is at most of the order of the interaction range, in the form of expansions in powers of a small parameter $\lambda/a$ \cite{hauteT} 
or $a/\lambda$ \cite{basseT}, where $\lambda$ is the
thermal de Broglie wavelength.

Interest in $b_3$ was reactivated by recent experimental breakthrough with cold atoms: 
long-lived spin $1/2$ Fermi gases can be prepared in the resonantly interacting regime
($|a|\!\gg\!$ interaction range) via Feshbach resonances \cite{manips_gen}. 
This motivated numerical calculation of $b_3$ in the maximally interacting, unitary limit $1/a\!=\!0$,
with the harmonic regulator technique of \cite{ComtetOuvry} as done in \cite{Drummond},
or with diagrams \cite{LeyronasKaplan}. Due to scaling invariance in the unitary limit,
$b_3$ is just a number, and via a precise measurement of the gas equation of state \cite{EOS,Salomon},
its predicted value was confirmed \cite{Salomon}.

Physics is richer when the Efimov effect \cite{Efimov} sets in: 
the continuous scaling invariance is broken, there appears a length scale $R_t$ characterizing the
interaction, the three-body parameter, and there exists an infinite number of trimer states
with an asymptotically geometric spectrum. The third cluster coefficient 
$b_3$ becomes a function of temperature. In a spinless bosonic gas with zero-range
interactions, it was determined analytically \cite{CastinWernerCan}. 
Within the three-body hard-core model that fixes $R_t$ \cite{vonStecher},
Quantum Monte Carlo simulations have confirmed this analytical prediction and have shown that the third order cluster expansion
can provide a good description of the gas 
down to the liquid-gas
transition \cite{KrauthPiatecki}, 
exemplifying its usefulness.

The problem is even more intriguing when a system parameter allows
to switch on-and-off the Efimov effect, as in the two-component Fermi gas with 
adjustable mass ratio. For two identical fermions and a distinguishable particle, there is an Efimov effect if 
the fermion-other particle mass ratio $\alpha$ exceeds $\alpha_c=13.6069\ldots$ 
\cite{Efimov,PetrovBraaten}.
Up to now, the calculation of $b_3$ is numerical and limited to $\alpha <\alpha_c$ \cite{Blume}. Strikingly
it predicts that $b_3$ has an infinite derivative at $\alpha=\alpha_c$. As $b_3$ is a coefficient in the
grand potential $\Omega$, this would imply a singular derivative of $\Omega$ as a function of $\alpha$,
i.e.\  a first order phase transition, subsisting at arbitrarily low phase space density, i.e.\
at temperatures $T$ arbitrarily higher than the Fermi temperature $T_F$, contrarily to common expectations
for phase transitions.  The present work determines $b_3$ analytically and solves this paradox.

\section{The cluster expansion}

We consider a mixture of two fully polarized fermionic species, with single particle masses $m_1$ and $m_2$,
with no intraspecies interaction and a purely $s$-wave interspecies interaction, of negligible
range and infinite scattering length (unitary limit).
At thermal equilibrium in a cubic box, the total pressure {$P$} admits in the thermodynamic limit the cluster expansion
\be
\frac{P \lambda_r^3}{k_B T} = \sum_{(n_1,n_2)\in \mathbb{N}^2} b_{n_1,n_2} z_1^{n_1} z_2^{n_2} 
\ee
where $z_i$ are fugacities $\exp(\beta \mu_i)$, $\lambda_r=[2\pi\hbar^2/(m_r k_B T)]^{1/2}$ is the thermal de Broglie wavelength
associated to the reduced mass $m_r=m_1 m_2/(m_1+m_2)$ and temperature $T$, $\beta=1/(k_B T)$, $\mu_i$ is the chemical potential
of species $i$, and $\mathbb{N}$ is the set of all non-negative integers. 

To determine the cluster coefficients $b_{n_1,n_2}$ one can use the harmonic regulator trick \cite{ComtetOuvry}: one rather assumes
that the system is at thermal equilibrium in an isotropic harmonic trap, with the same trap frequency $\omega$ for the two species,
and one considers the cluster expansion of $(-\Omega)/(k_B T Z_1)$ in powers of $z_1$ and $z_2$,
with $\Omega$ the grand potential and $Z_1$ the single particle partition function in the trap.
When $\omega\to 0$, the corresponding coefficients have a limit $B_{n_1,n_2}$ that one can relate
to $b_{n_1,n_2}$ \cite{ComtetOuvry,Drummond,Blume}:
\be
B_{n_1,n_2}=\left(\frac{m_r}{n_1 m_1+n_2 m_2}\right)^{3/2} b_{n_1,n_2} 
\ee
We study $B_{2,1}$ as a function of the mass ratio
$\alpha=m_1/m_2$.

\section{Case $\alpha<\alpha_c$: 0-parameter zero-range model}
\label{sec:0p0r}

The cluster coefficient $B_{2,1}$ can be deduced from the partition functions of up to three bodies in the trap,
that is from the $n$-body energy spectra for $n\leq 3$. In the unitary limit, the interspecies interaction 
is described by the Bethe-Peierls binary contact condition on the wavefunction, which leads to a separable
three-body
Schr\"odinger equation in internal hyperspherical coordinates \cite{Efimov} even in a harmonic trap \cite{WernerCastin,CastinWerner,WernerThese}.
The hyperangular part of the problem can be solved in position space \cite{Efimov}
or in momentum space \cite{MinlosFaddeev}: the corresponding real eigenvalue $s^2$ (that will serve as a separability constant)
obeys the transcendental equation $\Lambda_{l}(s)=0$ of explicit expression \cite{TignoneCastin}\footnote{There exists a less explicit hypergeometric expression for $\Lambda_l$ \cite{hypergeom}.}
\be
\Lambda_l(s)= \cos \nu + \frac{1}{\sin \nu} \int_{\frac{\pi}{2}-\nu}^{\frac{\pi}{2}+\nu}
\mathrm{d}\theta\, P_l\left(\frac{\cos\theta}{\sin \nu}\right) \frac{\sin (s\theta)}{\sin(s\pi)}
\ee
with $l\in\mathbb{N}$ the angular momentum, $P_l$ a Legendre polynomial, $\nu=\arcsin \frac{\alpha}{1+\alpha}$ the mass angle.
We call $(u_{l,n})_{n\in\mathbb{N}}$ the positive roots of $\Lambda_l$, sorted in increasing order. There is no complex root
for $\alpha< \alpha_c$.
The hyperradial part of the wavefunction, after multiplication by $R^2$, solves an effectively bidimensional Schr\"odinger equation:
\be
E F = -\frac{\hbar^2}{2M} \left(F''+\frac{1}{R} F'\right) + \left(\frac{\hbar^2 s^2}{2M R^2}+
\frac{1}{2} M \omega^2 R^2\right) F
\label{eq:ebse}
\ee
where $s$ is any of the $u_{l,n}$, $M=2 m_1+m_2$ is the mass of two particles of species 1 and one particle of species 2,
and the hyperradius $R$ is the corresponding mass-weighted root-mean-square deviation of the positions of the three particles from their center of mass.
Solving Eq.(\ref{eq:ebse}) with the usual boundary conditions that
$F(R)$ vanishes at zero and infinity gives
\be
E=(s+1+2q) \hbar \omega,  \ \ \ \forall q\in\mathbb{N}
\label{eq:univ_spectrum}
\ee
The semi-infinite ladder structure of this spectrum, with equidistance $2\hbar \omega$, reflects the existence of an undamped breathing
mode of the trapped non-Efimovian unitary gas \cite{CastinCRAS} related to its $SO(2,1)$ dynamical symmetry \cite{PitaevskiiRosch}.

\begin{figure}[tbp]
\begin{center}
\includegraphics[width=0.85\linewidth,clip=]{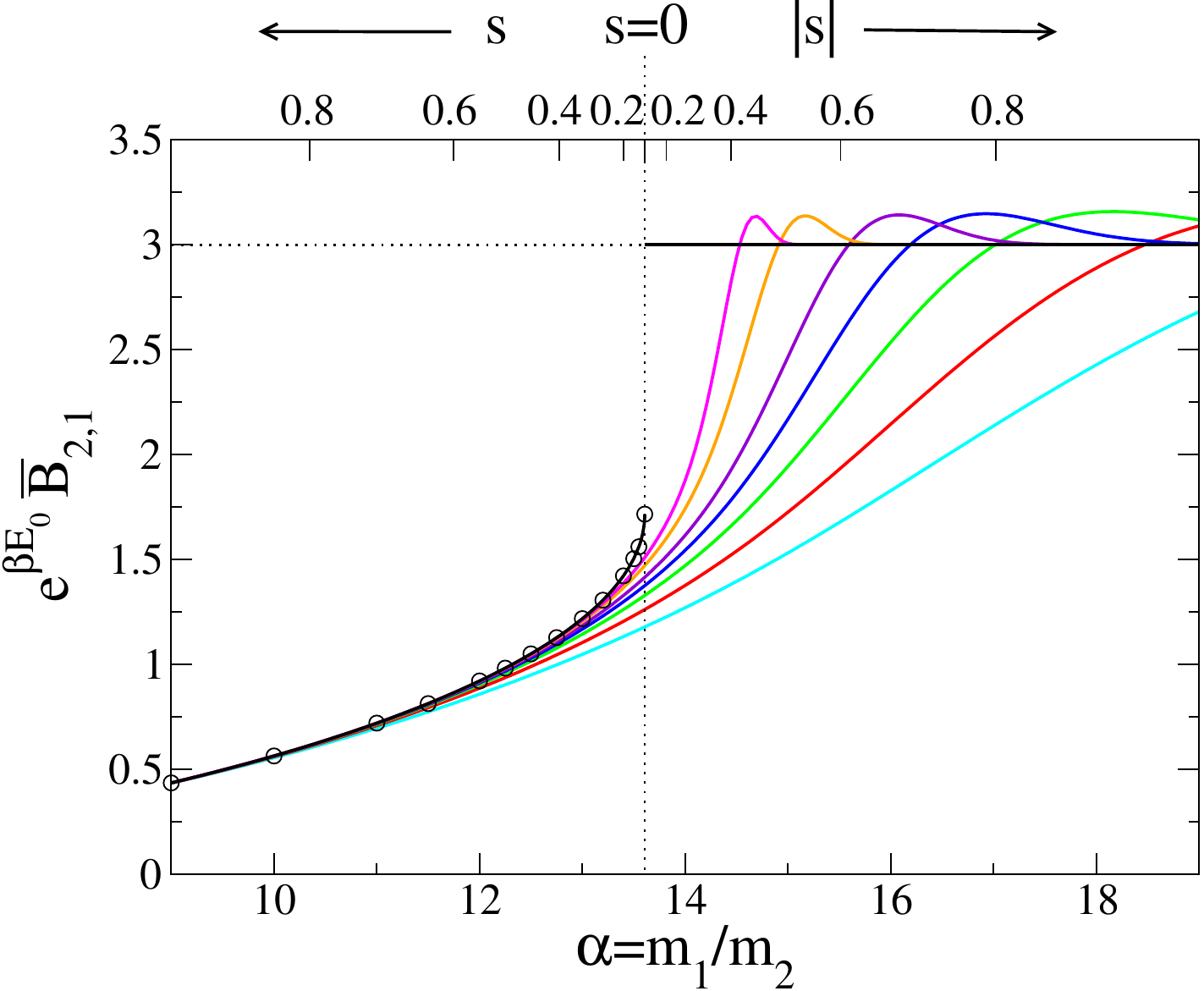}
\end{center}
\caption{Reduced third cluster coefficient $e^{\beta E_0} \bar{B}_{2,1}$ of a trapped two-component three-dimensional unitary Fermi gas
in the zero trapping frequency limit, as a function of the mass ratio $\alpha=m_1/m_2$ of the two species {(lower $x$-axis)
or of the root $s=u_{1,0}$ of $\Lambda_1$ (upper $x$-axis)},
for various values of the three-body parameter $R_t$, and hence of the global energy scale $E_{\rm glob}$ of
Eqs.(\ref{eq:eglob_usuel},\ref{eq:eglob_etendu}). Here $E_0$ is the ground free-space three-body energy, a smooth function of $\alpha$:
for $\alpha \leq \alpha_c$, $E_0=0$;  for $\alpha>\alpha_c$, $E_0=-E_{\rm glob} \exp(-2\pi/|s|)$ is the ground trimer energy 
and the factor $\exp(\beta E_0)$ ensures that the plotted quantity remains bounded.
Curves from bottom to top for $\alpha\lesssim 15$: $\beta E_{\rm glob}=10^2 \mbox{(cyan)}, 3\times 10^2 \mbox{(red)},
10^3 \mbox{(green)}, 3\times 10^3 \mbox{(blue)}, 10^4 \mbox{(violet)}, 10^5 \mbox{(orange)}, 10^6 \mbox{(magenta)}$.
The curves cross, which shows that
$e^{\beta E_0}\bar{B}_{2,1}$ is not, at all fixed $\alpha$, an increasing function of $\beta E_{\rm glob}$
{(see inset in Fig.~\ref{fig:sca})}. 
Discontinuous black solid line: limit $\beta E_{\rm glob}\to +\infty$, corresponding for $\alpha \leq \alpha_c$ 
to the genuine 0-parameter zero-range model studied numerically in \cite{Blume} (black circles),
and being, for $\alpha>\alpha_c$, identically equal to $3$, the ground-trimer contribution.
Vertical dotted line: critical mass ratio $\alpha_c$ where the Efimov effect sets in. 
} 
\label{fig:b21}
\end{figure}

Finally $B_{2,1}$ is the $\omega\to0$ limit of a series {\cite{Drummond,CastinWernerCan}}\footnote{Actually one calculates the difference between partition functions
of unitary and non-interacting problems; still this directly gives $B_{2,1}$ of the unitary gas 
since $B_{2,1}$ is zero for the ideal gases; the contributions of the
Laughlinian states (whose wavefunction vanishes when two particles are at the same point) 
cancel out in the difference; the $v_{l,n}$ appear via the non-Laughlinian spectrum of the non-interacting
three-body problem.  Similarly, the contributions of the unphysical root $s=2$ in the sector $l=0$, which exists in both the unitary and non-interacting
cases, automatically cancel out.}:
\be
B_{2,1}\!=\!\lim_{\omega\to 0}\!\!\!\!\!\!\sum_{(l,n,q)\in\mathbb{N}^3}\!\!\!\!\!\! (2l+1)\!\!
\left[e^{-(u_{l,n}+1+2q)\beta\hbar\omega}\!\!-\!e^{-(v_{l,n}+1+2q)\beta\hbar\omega}\right]
\label{eq:B21series}
\ee
with $v_{l,n}=l+2n+1$ the positive poles of $\Lambda_l(s)$ \cite{TignoneCastin}.
The summation over $q$ can be done, and even over $n$ by inverse application of the residue theorem \cite{CastinWernerCan}:
\be
B_{2,1}=-\sum_{l\in\mathbb{N}} \left(l+\frac{1}{2}\right) \int_0^{+\infty} \frac{\mathrm{d}S}{\pi} \ln\frac{\Lambda_l(iS)}{\cos\nu}
\label{eq:b21integ}
\ee
As shown in Fig.~\ref{fig:b21}, the result agrees with the numerical evaluation of the series by \cite{Blume}. The analytics however
directly allows to see why $B_{2,1}$ has an infinite derivative with respect to $\alpha$ at $\alpha=\alpha_c^-$: it suffices 
to isolate the contribution of the channel $(l,n)=(1,0)$ in Eq.(\ref{eq:B21series}), the only one where $u_{l,n}$ vanishes at $\alpha=\alpha_c$, by
the splitting
\be
B_{2,1}=B_{2,1}^{(1,0)}+B_{2,1}^{\neq(1,0)}
\ee
All the other channels have $u_{l,n}>1$ over the figure range and give a smooth contribution to $B_{2,1}$. 
On the contrary
\begin{eqnarray}
B_{2,1}^{(1,0)} &=&\lim_{\omega\to 0} 3\sum_{q\in\mathbb{N}}\! 
\left[e^{-(u_{1,0}+1+2q)\beta\hbar\omega}\!-\! e^{-(v_{1,0}+1+2q)\beta\hbar\omega}\right]
\nonumber \\
&=&-\frac{3}{2} (u_{1,0}-v_{1,0}) 
\label{eq:B21univ} 
\end{eqnarray}
and $u_{1,0}$, {a decreasing function of $\alpha$},
vanishes as $(\alpha_c-\alpha)^{1/2}$ since $\Lambda_1(s)$ is even, so that $\frac{\mathrm{d}}{\mathrm{d}\alpha} B_{2,1}$ diverges
as $(\alpha_c-\alpha)^{-1/2}$.

\section{Case $\alpha>\alpha_c$: Efimov zero-range model}
We now assume that the mass ratio obeys $\alpha_c<\alpha<75.99449\ldots$ \cite{Kartav},
so that the Efimov effect takes place in the sector $l=1$ only. The function $\Lambda_{l=1}$
has a pair of complex conjugate purely imaginary roots $\pm s$ and we set
\be
{u_{1,0}}=s=i |s|
\ee
{$|s|$ vanishes as $(\alpha-\alpha_c)^{1/2}$ and increases with $\alpha$.}
The $1/R^2$ potential in Schr\"odinger's equation (\ref{eq:ebse}) for $F(R)$ becomes attractive, which leads
to a ``fall to the center" \cite{Landau} and to an unphysical continuous spectrum of bound states, forcing to
modify the boundary condition at $R=0$ \cite{Morse}:
\be
F(R)\underset{R\to 0}{=} (R/R_t)^{i|s|}-(R/R_t)^{-i|s|}+O(R^2)
\label{eq:Morse}
\ee
To make evident that the third cluster coefficient now depends on {\sl one} parameter, 
this length $R_t$ called three-body parameter, we write it as $\bar{B}_{2,1}$, that is with {\sl one} overlining bar.
In free space, Eq.(\ref{eq:Morse}) leads to a discrete infinite number of Efimov trimer states, with a purely geometric
spectrum extending from $-\infty$ to $0$.
In any physical system, however, the interaction is not strictly zero range and the spectrum must be bounded from below \cite{Efimov}. 
One may expect that finite range effects then spoil the geometric nature of the spectrum for the more deeply bound trimers. 
However, for a narrow Feshbach resonance
\cite{TignoneCastin,Pricoupenko}, for momentum-space
cut-off models of a Feshbach resonance \cite{PricoupenkoJonaLasinio,EndoNaidonUeda}, and for the three-body hard core model \cite{vonStecher}, 
the spectrum is almost entirely geometric, at least when $|s|$ is not too large ($|s|\lesssim 1)$, and becomes
entirely geometric when $\alpha\to \alpha_c^+$, since the typical particle wavenumber times the interaction range tends to zero
\cite{TignoneCastin}. In what follows, we assume the bounded from below geometric free space spectrum:
\be
\epsilon_q(0^+){\equiv \lim_{\omega\to 0} \epsilon_q(\omega)}
=-E_{\rm glob} e^{-2\pi(1+q)/|s|},\ \ \forall q\in\mathbb{N}
\label{eq:fss}
\ee
The global energy scale $E_{\rm glob}$ can be calculated from a microscopic model for the interaction,
as it was done in the above mentioned models. Here we take it as a parameter that solution of
Eq.(\ref{eq:ebse}) with $\omega=0$ and with the boundary condition (\ref{eq:Morse}) relates to $R_t$ as
\be
E_{\rm glob}= \frac{2\hbar^2}{MR_t^2} e^{[\ln \Gamma(1+s)-\ln \Gamma(1-s)]/s}
\label{eq:eglob_usuel}
\ee
with $\ln\Gamma$ the usual branch of the $\Gamma$ function logarithm.

The contribution to $\bar{B}_{2,1}$ of the channels $(l,n)\neq (1,0)$ is unchanged since no Efimov effect occurs in these channels:
\be
\bar{B}_{2,1}^{\neq(1,0)} = B_{2,1}^{\neq(1,0)}
\ee
We calculate it as in \cite{CastinWernerCan}, using Eq.(\ref{eq:b21integ}) as it is for $l\neq 1$, while substituting
$\Lambda_l(iS)$ with $\frac{S^2+v_{1,0}^2}{S^2+u_{1,0}^2}\Lambda_l(iS)$ for $l=1$.  
In the Efimovian channel $(l,n)=(1,0)$, the spectrum is no longer given by Eq.(\ref{eq:univ_spectrum}), 
but by the solution of the transcendental equation deduced from \cite{Pethick} and rewritten as in \cite{CastinWernerCan,Gen2}
to match Eq.(\ref{eq:fss}) in free space:
\be
\im \ln \Gamma\Big(\frac{1+s-\epsilon_q/(\hbar\omega)}{2}\Big)
+\frac{|s|}{2} \ln \Big(\frac{2\hbar\omega}{E_{\rm glob}}\Big) + (q+1) \pi =0
\ee
so that the first identity in Eq.(\ref{eq:B21univ}) is replaced by
\be
\bar{B}_{2,1}^{(1,0)}=\lim_{\omega\to 0} 3\sum_{q\in\mathbb{N}} \left[e^{-\beta\epsilon_q(\omega)} - e^{-(v_{1,0}+1+2q)\beta \hbar\omega}
\right]
\ee
For a small enough non-zero $\omega$, two classes emerge in the three-body spectrum: 
(i) negative {eigenenergies}, that are the equivalent of free space trimer {energies}, 
and (ii) positive {eigenenergies}, that are the equivalent of the free space continuum. The second class 
{is} a harmonic spectrum except for an energy dependent ``quantum defect" 
$\Delta(\epsilon)$ \cite{Gen2}
\be
\frac{\epsilon_q(\omega)}{\hbar\omega}\underset{q\to+\infty}{=} 2q + \Delta(\epsilon_q(\omega)) + O(1/q)
\ee
where $q\omega$ is $\approx$ fixed. By the reasoning of \cite{CastinWernerCan}, we get
\begin{multline}
\bar{B}_{2,1}^{(1,0)}= 3\sum_{q\in\mathbb{N}} [e^{-\beta \epsilon_q(0^+)}-1] \\
-\frac{3}{2} \int_0^{+\infty} \mathrm{d}\epsilon \beta [\Delta(\epsilon)-(1+v_{1,0})] e^{-\beta \epsilon}
\label{eq:b2110efim}
\end{multline}
We obtained a new expression of the quantum defect\footnote{This expression and the one (C6) of \cite{Gen2} are equal, since
their difference is a continuous function of $x$ that vanishes at zero and has an
identically zero derivative.}:
\be
\Delta(\epsilon)=2+\frac{2}{\pi} \atan \frac{\tan (\frac{|s|}{2}x )}{\tanhyper (\frac{|s|}{2}\pi)}
+2\left\lfloor \frac{|s|x}{2\pi}\right\rceil
\label{eq:Delta_efim}
\ee
where $x=\ln(\epsilon/E_{\rm glob})$. The nearest-integer function in the last term exactly compensates the jumps of the 
$\atan$ function when $\tan(|s|x/2)$ diverges, so as to render $\Delta(\epsilon)$ a smooth function of $\epsilon$ and
of $|s|$.

The corresponding values of $\bar{B}_{2,1}$ for $\alpha>\alpha_c$ are shown in Fig.\ref{fig:b21}, after multiplication
by a factor $e^{\beta E_0}$, where $E_0=\epsilon_{q=0}(0^+)$ is the ground trimer energy, so as to absorb its contribution 
that becomes rapidly dominant and divergent for $k_B T < |E_0|$ \cite{PaisUlhenbeck1959}.
The result depends on $\beta E_{\rm glob}$, a parameter that must be $\gg 1$:
our theory, being zero range, requires that $R_t$, of order of the interaction range or effective range, as in the three-body hard core
and narrow Feshbach resonance models respectively, is $\ll$ the thermal de Broglie wavelength ${\lambda}_t=[2\pi\hbar^2/(M k_B T)]^{1/2}$.
Clearly, there is a discrepancy of $B_{2,1}(\alpha)$ and $\bar{B}_{2,1}(\alpha)$ at $\alpha_c^{\mp}$ at non-zero $R_t$. When $R_t\to 0$
($E_{\rm glob}\to +\infty$) there is agreement at $\alpha_c$, as seen by first taking the $s\to 0$ limit in Eq.(\ref{eq:Delta_efim}),
\be
\Delta(\epsilon)\underset{\alpha\to\alpha_c^+}{\to} \Delta_0(\epsilon)=2+\frac{2}{\pi} \atan \frac{\ln(\epsilon/E_{\rm glob})}{\pi}
\label{eq:Delta0}
\ee
then taking the $R_t\to 0$ limit in Eq.(\ref{eq:b2110efim})\footnote{One takes $\beta \epsilon$ as integration variable and one expands the integrand in powers of $1/\ln(\beta E_{\rm glob})$.}:
\begin{multline}
\bar{B}_{2,1}^{(1,0)}(\alpha_c^+) = -\frac{3}{2} \int_0^{+\infty} \!\!\!\!\!\!\!\! \mathrm{d}\epsilon \beta [\Delta_0(\epsilon)-(1+v_{1,0})] e^{-\beta \epsilon} \\
\underset{\beta E_{\rm glob}\to +\infty}{=} \frac{3}{2} v_{1,0} - \frac{3}{\ln(\beta E_{\rm glob})} + O\left(\frac{1}{\ln (\beta E_{\rm glob})}\right)^2 
\end{multline}
successfully collated with the $u_{1,0}\to 0$ value of Eq.(\ref{eq:B21univ}).
The key point however is that this $R_t\to 0$ limit is in practice inaccessible, due to the very slow logarithmic convergence.
We expect this problem to extend to $\alpha<\alpha_c$, which makes the strictly zero-range calculation of \cite{Blume} 
not fully realistic.
There also remains the puzzle of the diverging derivative of $B_{2,1}(\alpha)$ with respect to $\alpha$ at $\alpha_c^-$.
Both issues are solved in the next section.

\section{Case $\alpha < \alpha_c$ revised: 1-parameter zero-range model}

We  now see that a three-body parameter $R_t$ must be introduced for $\alpha<\alpha_c$, i.e.\ even in the absence
of Efimov effect, when $\alpha$ is close enough to $\alpha_c$. The root $s=u_{1,0}>0$ then {vanishes
as $(\alpha_c-\alpha)^{1/2}$} and the centrifugal barrier 
in the hyperradial equation (\ref{eq:ebse}) weakens, so that the function $F(R)$, the eigenenergies
$E$ and the third cluster coefficient become increasingly sensitive to short distance physics of the interaction \cite{EndoNaidonUeda,BlumeCond}.

Assume that three-body physics inside the interaction range is described by an extra term $V(R) F$ 
compared to Eq.(\ref{eq:ebse}),
e.g.\ a three-body hard core of radius $b$. Knowing that the relevant eigenenergies $E$ are at most a few $k_B T$, 
and that $b \ll \lambda_t$, we can make the following reasonings.

(i) at $R \ll{\lambda}_t$, one can obtain the behavior of $F(R)$ by a zero-energy calculation (neglecting the $EF$ term) in
free space (since the harmonic oscillator length is $\gg{\lambda}_t$). Due to $b \ll {\lambda}_t$
there exists a range $b\ll R\ll {\lambda}_t$ where one can also neglect $V(R)$. Then $F(R)$ is a superposition of the two
particular solutions $R^s$ and $R^{-s}$, with relative amplitudes fixed by {a length $R_t$ that depends on}
microscopic details of $V(R)$, e.g.\ $R_t=b$ for the three-body hard core\footnote{
If one sets $F(R)=R^{-s} \phi(r=R^{2s})$ then $F''+F'/R-s^2 F/R^2=4s^2R^{3s-2} \phi''(r)$ so
 that $R_t^{2s}=a_{\rm eff}$, where $a_{\rm eff}$ is the $s$-wave scattering ``length" of a particle of mass $M$
on the potential $v(r)=V(r^{1/(2s)})r^{-2+1/s}/(4s^2)$.  We suppose here that $a_{\rm eff}>0$, e.g.\ because $V(R)$ is non-negative.}:
\be
F(R) \underset{b\ll R \ll {\lambda}_t}{\simeq} (R/R_t)^s -(R/R_t)^{-s}
\label{eq:Fsd}
\ee

(ii) one can approach the same range $b\ll R\ll {\lambda}_t$ from large distances. 
The trapping potential and the $EF$ term must now be kept, and $F(R)$ is the unique
solution (up to normalisation) of Eq.(\ref{eq:ebse}) that does not diverge at infinity, 
a Whittaker function {of $R^2$} divided by $R$ \cite{WernerThese}.
Then at $R \ll{\lambda}_t$, $F(R)$ is also found to be a linear superposition of $R^s$ and $R^{-s}$, as it must be, 
but with coefficients $A_\pm(E)$ that are known functions of $E$. Matching with Eq.(\ref{eq:Fsd}) gives an implicit equation 
for $E$, as if Eq.(\ref{eq:ebse}) was subjected to the modified boundary condition at $R=0$ 
\cite{CastinWerner,NishidaTan}\footnote{For $s=0$ this becomes
$F(R)\underset{R\to0}{=}\ln(R/R_t)+O(R^2\ln R)$.}:
\be
F(R)\underset{R\to 0}{=} (R/R_t)^s-(R/R_t)^{-s} + O(R^{2-s})
\label{eq:bcm}
\ee
The third term in Eq.(\ref{eq:bcm}), coming from a property of the Whittaker function, is negligible as compared to the first one, and
this model makes sense, for $s<1$ {$\mathrm{i.e.}\ \alpha > 8.6185\ldots$}.
Remarkably this reproduces the Efimov zero-range model (\ref{eq:Morse}) if one formally replaces $s$ by $i |s|$. Then it is natural to extend
the definition of $E_{\rm glob}$ to $\alpha<\alpha_c$
\footnote{On a narrow resonance of Feshbach length $R_*$ one gets from \cite{TignoneCastin}
$(\frac{m_r R_*^2}{2\hbar^2} E_{\rm glob})^s=\frac{1-s}{1+s}\frac{\Gamma(1+2s)}{\Gamma(1-2s)} f(v_{1,0})
\prod_{n\in\mathbb{N}^*} \frac{f(v_{1,n})}{f(u_{1,n})}$ with $f(z)={\Gamma(z-s)\Gamma(1+z-s)}/{[\Gamma(z+s)\Gamma(1+z+s)}]$.}:
\be
E_{\rm glob} \underset{0<s<1}{=}\left(\frac{\Gamma(1+s)}{\Gamma(1-s)}\right)^{1/s} \frac{2\hbar^2}{MR_t^2}
\label{eq:eglob_etendu}
\ee
where the first factor is a smooth function of $\alpha$, as its series expansion involves only even powers of $s$.

The more common boundary condition $F(R=0)=0$, that led to the spectrum (\ref{eq:univ_spectrum}), is 
usually justified as follows: at $R\approx {\lambda}_t$, the $R^{-s}$ term in (\ref{eq:Fsd}) is negligible 
as compared to the $R^s$ term in the zero-range limit $b\ll {\lambda}_t$, that is $\beta E_{\rm glob}\gg 1$ as one expects $R_t \approx b$
\footnote{In peculiar cases, known as three-body resonances, see \cite{CastinWerner,NishidaTan},
$R_t/b$ can be arbitrarily large and $\beta E_{\rm glob}$ can remain finite in the zero-range limit. 
This is improbable here as there is already a two-body resonance.}:
\be
\frac{({\lambda}_t/R_t)^{-s}}{({\lambda}_t/R_t)^{s}} \approx (\beta E_{\rm glob})^{-s} \ll 1
\ee
However this condition becomes more and more difficult to satisfy when $\alpha\to \alpha_c^-$, and it will be violated when
\be
s \lesssim \frac{1}{\ln (\beta E_{\rm glob})}
\label{eq:transintui}
\ee
This forces us to recalculate the third cluster coefficient with the boundary condition (\ref{eq:bcm}).
From the implicit equation for the energy spectrum $(\epsilon_q(\omega))_{q\in \mathbb{N}}$ \cite{WernerThese} 
\footnote{The ground state solution
of this equation must be omitted, because it connects when $\omega\to 0$ to a bound state
of energy $-E_{\rm glob}$ and spatial extension $\approx R_t$, which cannot be faithfully described by our zero-range model
when $R_t\approx b$ (i.e.\ in the absence of three-body resonance) and indeed does not exist in the three-body hard-core
or in the narrow Feshbach resonance model \cite{TignoneCastin}. 
{This is equivalent to the assumption in \cite{Blume} of the absence of non-universal trimer states.}
}:
\be
\frac{\Gamma(\frac{1+s-E/\hbar\omega}{2})}{\Gamma(\frac{1-s-E/\hbar\omega}{2})} =
\left(\frac{E_{\rm glob}}{2\hbar\omega}\right)^s
\ee
we recalculate the quantum defect as in \cite{Gen2}, using the Euler reflection and Stirling formulas:
\be
\Delta(\epsilon) = 2 + \frac{2}{\pi} \atan \frac{\tanhyper [\frac{s}{2} \ln (\epsilon/E_{\rm glob})]}{\tan (\frac{s}{2}\pi)}
\label{eq:Delta_revis}
\ee
When $R_t\to 0$, $\beta E_{\rm glob}\to +\infty$ and this reproduces the value $1+s$ of the quantum defect 
in Eq.(\ref{eq:univ_spectrum}).  
Eq.(\ref{eq:Delta_revis}) only revises the contribution of the channel $(1,0)$,
since the other channels have $u_{l,n}>1$ for the values of $\alpha$ in Fig.~\ref{fig:b21}:
\be
\bar{B}_{2,1}^{(1,0)}\underset{0<s<1}{=}-\frac{3}{2} \int_0^{+\infty} \mathrm{d}\epsilon \beta [\Delta(\epsilon)-(1+v_{1,0})] e^{-\beta\epsilon}
\label{eq:B21_revis}
\ee
In Fig.~\ref{fig:b21} we plot for $\alpha<\alpha_c$ the corresponding values of $\bar{B}_{2,1}$, for the same values of the parameter
$\beta E_{\rm glob}$ as in the part $\alpha>\alpha_c$ of the figure, leading to an apparently smooth connection at $\alpha=\alpha_c$.
The continuity of the connection could be expected from the fact that (i) the formal change $s\to i|s|$ in Eq.(\ref{eq:Delta_revis}) 
reproduces the value (\ref{eq:Delta_efim}) of the quantum defect on the side $\alpha>\alpha_c$ apart from the nearest-integer function 
which is irrelevant when $|s|\to 0$, 
and (ii) the Efimovian trimer spectrum has a vanishing contribution to $\bar{B}_{2,1}$ when $\alpha\to \alpha_c^+$.

Indeed $\bar{B}_{2,1}^{(1,0)}(\alpha)$ (and $\bar{B}_{2,1}(\alpha)$) are smooth functions of $\alpha$ at $\alpha_c$ at fixed $\beta E_{\rm glob}$,
since $\Delta(\epsilon)$ is an even function of $s$ and its series expansion only has even powers of $s$:
\be
\Delta(\epsilon)\underset{s\to 0}{=}\Delta_0(\epsilon)-\frac{x}{6} s^2 +\frac{x^3-\pi^2 x}{360} s^4 + O(s^6)
\ee
where $\Delta_0(\epsilon)$ is given by Eq.(\ref{eq:Delta0}), $x=\ln(\epsilon/E_{\rm glob})$, and $s$ can be real or purely imaginary. 
Insertion in Eq.(\ref{eq:B21_revis}) leads to converging integrals over $\epsilon$ and to an expansion of $\bar{B}_{2,1}^{(1,0)}$ 
with only even powers of $s$:
\begin{multline}
\bar{B}_{2,1}^{(1,0)}(\alpha)-\bar{B}_{2,1}^{(1,0)}(\alpha_c) \underset{s\to 0}{=} -\frac{A}{4} s^2 \\
-\frac{A(\pi^2-{2}A^2)-4\zeta(3)}{480}s^4 + O(s^6)
\label{eq:dev}
\end{multline}
where $A=\ln(e^\gamma\beta E_{\rm glob})$ and $\gamma\simeq 0.577$ is Euler's constant
\footnote{Exchange of Taylor 
expansion and integration is justified by the theorem of derivation
under the integral, where $x$ is the integration variable.
For $\alpha<\alpha_c$, one sets $u(x,s)=\tanhyper(sx/2)/s$ and $v(x,s)=\tan(s\pi/2)/s$ and one fixes some $\eta\in ]0,1[$.
Then there exist positive numbers $(A_n,B_n)_{n\in\mathbb{N}}$ and  $C>0$ such that 
$\forall (x,s)\in\mathbb{R}\times[0,\eta]$, $\forall n\in \mathbb{N}$:
$|\partial_s^n u(x,s)|\leq A_n |x|^{n+1}$, $|\partial_s^n v(x,s)|\leq B_n$, $u(x,s)^2+v(x,s)^2\geq C$.
For $\alpha>\alpha_c$ one sets $u(x,|s|)=[\pi\sin(|s|x)-x\sinh(\pi|s|)]/|s|^2$ and $v(x,|s|)=[\cosh(\pi|s|)
-\cos(|s|x)]/|s|^2$ and one fixes some $\eta>0$. Then there exist positive numbers $(A_n,B_n,C_n,D_n)_{n\in\mathbb{N}}$ and $G>0$ such that
$\forall (x,|s|)\in\mathbb{R}\times[0,\eta]$, $\forall n\in \mathbb{N}$:
$|\partial_{|s|}^n u(x,|s|)|\leq A_n |x| +B_n |x|^{n+2}$, $|\partial_{|s|}^n v(x,|s|)|\leq C_n + D_n |x|^{2+n}$,
$|v(x,|s|)|\geq G$. The $u$ and $v$ functions appear in $\partial_s \Delta(\epsilon)$ as $(v\partial_s u-u\partial_s v)/(u^2+v^2)$
for $\alpha<\alpha_c$, and in $\partial_{|s|} \Delta(\epsilon)$ as $u/v$ for $\alpha>\alpha_c$.
Then $|\partial_s^n\Delta|$ and $|\partial_{|s|}^n\Delta|$ are polynomially bounded in $|x|$ uniformly
in $s$ or $|s|$ $\forall n\in \mathbf{N}^*$.}.
Since $s^2$ is a smooth function of $\alpha$ across $\alpha_c$, so is $\bar{B}_{2,1}$.

\begin{figure}[tbp]
\begin{center}
\includegraphics[width=0.85\linewidth,clip=]{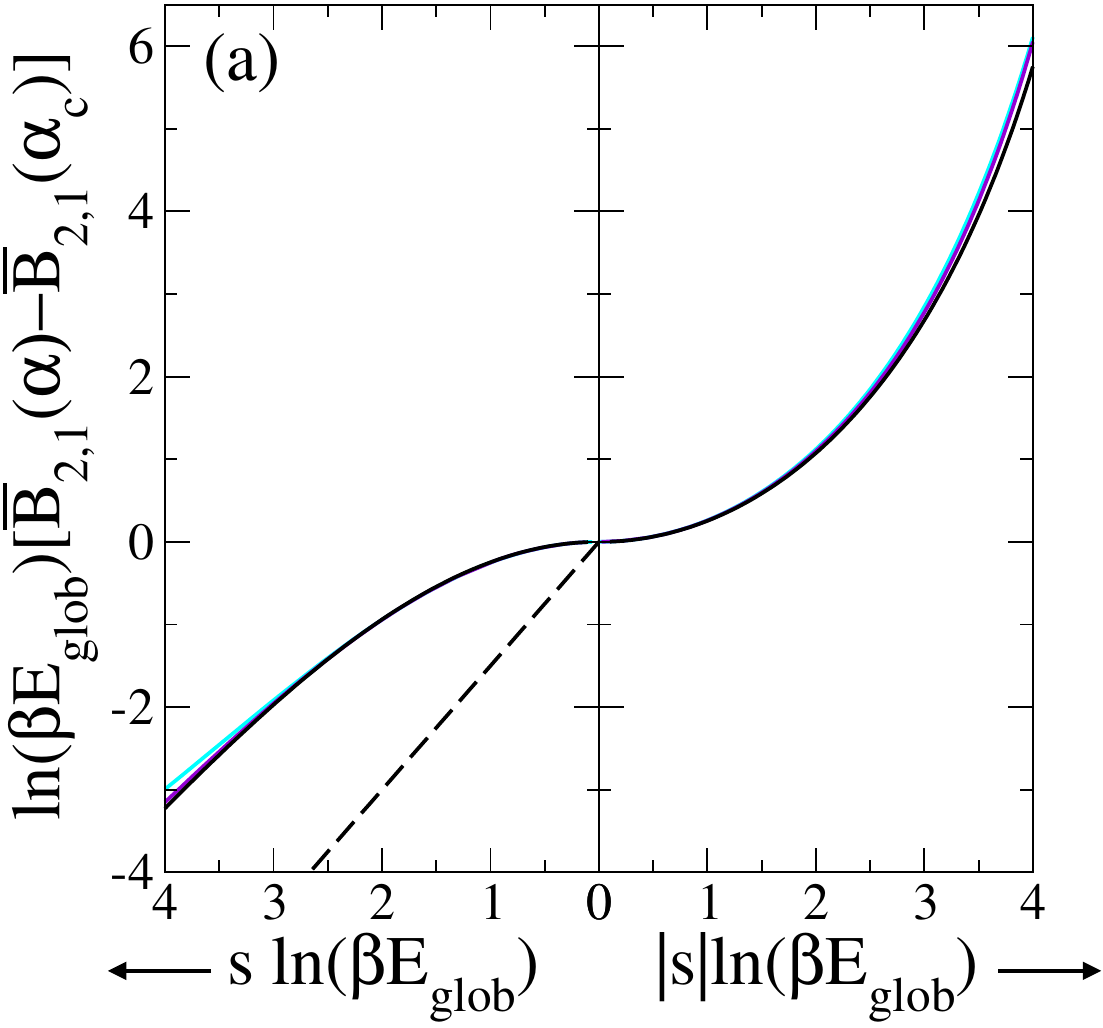} \\ \vspace{10mm} 
\includegraphics[width=0.85\linewidth,clip=]{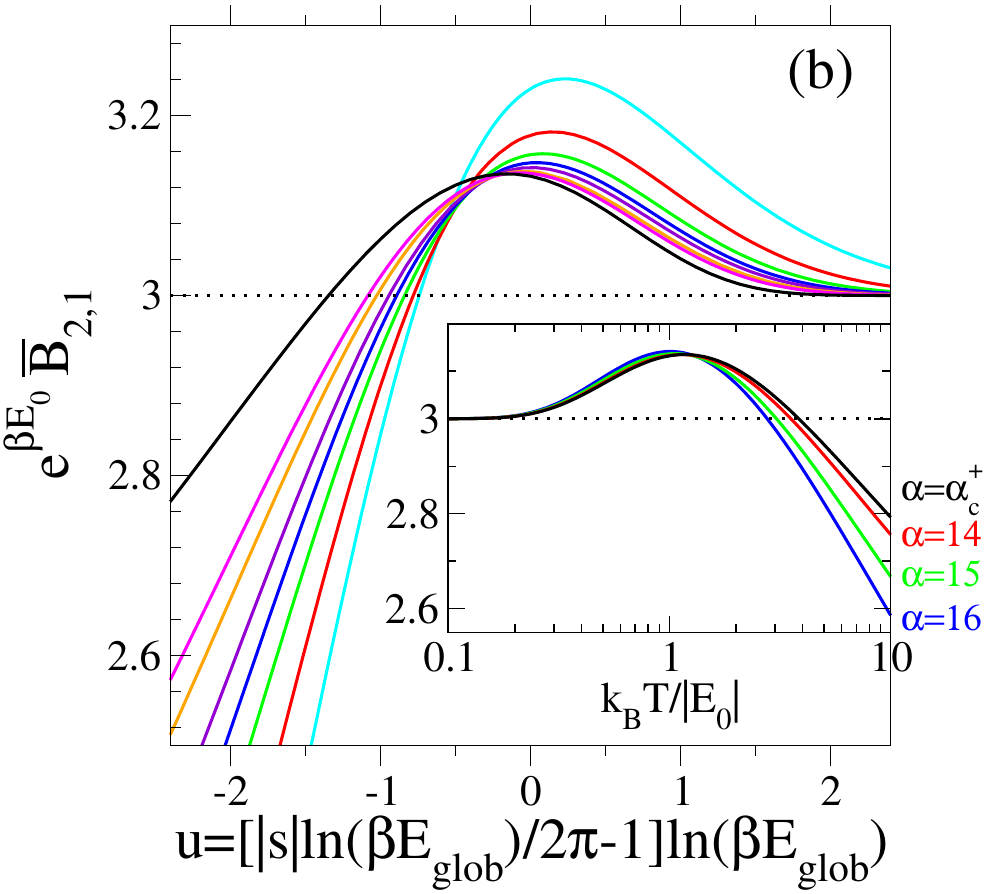}
\end{center}
\caption{Scaling-law analysis of Fig.\ref{fig:b21} for $\alpha$ close to $\alpha_c$ in the limit $\beta E_{\rm glob}\to +\infty$.
(a) For a fixed $t\equiv s \ln(\beta E_{\rm glob})$: the data approach the law (\ref{eq:scaling_law}) (black solid line), 
provided that $|t|<2\pi$ on the 
side $\alpha>\alpha_c$. Curves from top to bottom: $\beta E_{\rm glob}=10^2 \mbox{(cyan)}, 10^4 \mbox{(violet)}$. 
Dashed line: 0-parameter theory prediction $-3t/2$.
(b) On the side $\alpha>\alpha_c$, for a fixed $u\equiv [|t|/(2\pi)-1] \ln (\beta E_{\rm glob})$: the data approach the law (\ref{eq:scaling_law2})
(black solid line),
which {reproduces} the peaked structure seen in Fig.\ref{fig:b21}. From bottom to top for $u<-0.6$: values of $\beta E_{\rm glob}$ listed in
the caption of Fig.\ref{fig:b21}, with the same order and colors. 
{Inset: from Eqs.(\ref{eq:B21univ},\ref{eq:b2110efim},\ref{eq:Delta_efim}),
Eq.(\ref{eq:scaling_law2}) taken with $u=\ln(\beta|E_0|)$ also gives the limit of $e^{\beta E_0} \bar{B}_{2,1}$ 
for $\alpha\to \alpha_c^{+}$ at fixed $k_B T/|E_0|$.}
}
\label{fig:sca}
\end{figure}

Eq.(\ref{eq:dev}), combined with $\Lambda_1(s)=0$, predicts
how the first order derivative at $\alpha_c$ diverges when $\beta E_{\rm glob}\to +\infty$:
\be
\frac{\mathrm{d}}{\mathrm{d}\alpha} \bar{B}_{2,1}(\alpha_c)\!\!\!\underset{\beta E_{\rm glob}\to+\infty}{\sim}\!\!\! C \ln (\beta E_{\rm glob})
\ \mbox{with}\ C\simeq 0.0478243
\ee
It also suggests an interesting scaling law close to $\alpha_c$: keeping in the coefficients of the powers of $s$ in (\ref{eq:dev})
only the leading terms in $\ln(\beta E_{\rm glob})$, one
uncovers, after multiplication of (\ref{eq:dev}) by  $\ln(\beta E_{\rm glob})$, the following law when $\beta E_{\rm glob}$
tends to infinity at fixed $t\equiv s \ln (\beta E_{\rm glob})$:
\be
[\bar{B}_{2,1}(\alpha)-\bar{B}_{2,1}(\alpha_c)] 
\ln (\beta E_{\rm glob}) 
\stackrel{t\ \mathrm{fixed}}{\underset{\beta E_{\rm glob}\to +\infty}{\to}} 
3-\frac{3t/2}{\tanhyper(t/2)}
\label{eq:scaling_law}
\ee  
with no constraint on the side $\alpha<\alpha_c$, and with the constraint that 
$|t|<2\pi$ on the side $\alpha>\alpha_c$ due to the occurrence of a pole at $t=2\pi i$ in the quantum defect contribution and to a divergence of the ground trimer contribution for $|t|>2\pi$.
Eq.(\ref{eq:scaling_law}) is obtained by  neglecting $\ln(\beta \epsilon)$ as compared to $\ln(\beta E_{\rm glob})$ 
in (\ref{eq:Delta_efim},\ref{eq:Delta_revis}), as $\beta \epsilon$
is typically unity in the {integrals (\ref{eq:b2110efim},\ref{eq:B21_revis})}. 
In Fig.~\ref{fig:sca}a we replot the data of Fig.\ref{fig:b21} after rescaling as in Eq.(\ref{eq:scaling_law}):
the results are indeed almost aligned on a single scaling curve given by Eq.(\ref{eq:scaling_law}), the better the larger $\ln(\beta E_{\rm glob})$ is.
The 0-parameter zero-range theory prediction $-3t/2$, see dashed line, is only asymptotically
equivalent to the correct law at $t\to +\infty$. The scaling law fully justifies the intuitive condition (\ref{eq:transintui}):
the crossover from the 0- to the 1-parameter zero-range regime indeed occurs for $s\approx 1/\ln(\beta E_{\rm glob})$.

What happens on the side $\alpha>\alpha_c$ close to $|t|=2\pi$? 
For $|t|$ fixed to a value $>2\pi$, the ground-trimer contribution $3e^{-\beta E_0}$, where $E_0=\epsilon_0(0^+)$,
rapidly diverges when $\beta E_{\rm glob}\to +\infty$
and dominates all other contributions, so that the reduced cluster coefficient $e^{\beta E_0} \bar{B}_{2,1}$ of Fig.~\ref{fig:b21}
tends to three. However, before that, the reduced cluster coefficient exhibits as 
a function of $\alpha$ an interesting structure in 
Fig.~\ref{fig:b21}, a sharp rise with a maximum,
that corresponds to a neighbourhood of $|t|=2\pi$ with a width $1/\ln (\beta E_{\rm glob})$.  This is revealed by the affine change of variable
$u\equiv [|t|/(2\pi)-1] (\ln \beta E_{\rm glob})$. When $\beta E_{\rm glob}\to +\infty$ for fixed $u$, $\beta E_0\to -e^u$, the ground-trimer contribution
remains finite and, from dominated convergence theorem,
\begin{multline}
e^{\beta E_0} \bar{B}_{2,1} \stackrel{u\ \mathrm{fixed}}{\underset{\beta E_{\rm glob}\to +\infty}{\to}}
e^{-e^u}\left[B_{2,1}(\alpha_c)+3\left(e^{e^u}-\frac{1}{2}\right)\right. \\
\left. +\frac{3}{\pi} \int_0^{+\infty} \mathrm{d}\epsilon \beta e^{-\beta \epsilon} \atan \frac{u-\ln \beta\epsilon}{\pi}\right]
\label{eq:scaling_law2}
\end{multline}
where $B_{2,1}(\alpha_c)\simeq 1.7153$ \cite{Blume} is the prediction of the 0-parameter zero-range theory at $\alpha_c$.
As shown in Fig.~\ref{fig:sca}b, the rescaled data of Fig.~\ref{fig:b21} nicely converge to this law.

\section{Conclusion}
As compared to the usual zero-range theory
we have found corrections of order $1/\ln({\lambda}_t/R_t)$ to the third virial coefficient
of a two-component unitary Fermi gas, close to and below the threshold for the Efimov effect, at
a distance $\alpha_c-\alpha$ scaling as $1/[\ln({\lambda}_t/R_t)]^2$, where $R_t$ is a three-body parameter and
$\lambda_t$ a thermal de Broglie wavelength;
these $1/\ln({\lambda}_t/R_t)$ corrections {arise from short-range three-body correlations, that is from triplets of close atoms}\footnote{
For $s<1/2$ they dominate over more usual (here neglected) energy corrections, due to the effect of a finite range $b$ of
the two-body interaction
{at the level of short-range two-body correlations (only involving {\sl pairs} of close atoms)}, that vanish linearly in $b$ \cite{Gen1}.
From Eq.(\ref{eq:B21_revis}) for $R_t/\lambda_t\to 0$, $\bar{B}_{2,1}-B_{2,1}\sim
-\frac{3}{\pi}{\Gamma(s+1)\sin (s\pi)}/{(\beta E_{\rm glob})^s}$ then indeed vanishes more slowly than $b$, assuming that $R_t\approx b$.}.
As a consequence, for a given finite ${\lambda}_t/R_t$ as in all realistic systems, the third virial coefficient
reconnects smoothly to its values deduced from the Efimov zero-range model above 
the threshold, precluding the unphysical first-order phase transition predicted by zero-range theory. 
{Our predictions may be tested by measuring the equation of state of mixtures of fermionic cold atoms
with a mass ratio $\alpha\simeq 13.6$, such as ${}^3$He${}^*$ and ${}^{40}$K.}

\section{Acknowledgments} S.E. thanks JSPS for support.

\end{document}